\documentstyle[a4,12pt,amssymb,epsfig,times]{article}
\newcommand{\be}{\begin{equation}}
\newcommand{\ee}{\end{equation}}
\newcommand{\bea}{\begin{eqnarray}}
\newcommand{\eea}{\end{eqnarray}}

\def\Nu{{\cal{V}}}

\def\p1{\pi_1}
\def\a{\alpha}
\def\b{\beta}
\def\d{\delta}
\def\l{\lambda}
\def\f{\phi}
\def\r{\rho}

\def\ep{\epsilon}

\def\g{\gamma}
\def\z{\zeta}

\renewcommand{\thefootnote}{\fnsymbol{footnote}}
\renewcommand{\theequation}{\thesection.\arabic{equation}}
\begin{document}

\begin{titlepage}
\vspace*{\stretch{0}}
\begin{flushright}
{\tt FTUV/99-73\\
     IFIC/99-77\\
     hep-th/9911091}
\end{flushright}

\vspace*{0.5cm}

\begin{center}
{\Large\bf Holography, degenerate horizons and entropy}
\\[0.5cm]
Diego J. Navarro\footnote{\tt dnavarro@ific.uv.es},
Jos\'e Navarro-Salas\footnote{\tt jnavarro@lie.uv.es} and Pedro
Navarro\footnote{\tt pnavarro@lie.uv.es}
\\[0.5cm]
{\footnotesize
       Departamento de F\'{\i}sica Te\'orica and
       IFIC, Centro Mixto Universidad de Valencia-CSIC.\\
       Facultad de F\'{\i}sica, Universidad de Valencia,
       Burjassot-46100, Valencia, Spain.}
\end{center}
\bigskip
\begin{abstract}
We show that a realization of the correspondence AdS$_2$/CFT$_1$ for
near extre\-mal Reissner-Nordstr\"om black holes in arbitrary dimensional
Einstein-Maxwell gravity exactly reproduces, via Cardy's formula, the deviation
of the Bekenstein-Hawking entropy from extremality. We also show that this
mechanism is valid for Schwarzschild-de Sitter black holes around the degenerate
solution dS$_2\times$S$^n$. These results reinforce the idea that the
Bekenstein-Hawking entropy can be derived from symmetry principles.
\end{abstract}
\vspace*{\stretch{1}}
\begin{flushleft}
PACS number(s): 04.60.Kz; 04.60.Ds
\end{flushleft}
\vspace*{\stretch{0}}
\end{titlepage}
\newpage

\renewcommand{\thefootnote}{\arabic{footnote}}
\setcounter{footnote}{0}

\section{Introduction}

For a long time it has been a remarkable puzzle to unravel the origin
of the Bekenstein-Hawking entropy of a black hole. One would expect
that string theory, as a theory of quantum gravity, could offer a 
microscopic explanation of black hole entropy. However, only recently
it has been possible, by applying D-brane techniques, to perform precise 
calculations which succeed in reproducing the Bekenstein-Hawking entropy
for extremal \cite{sv, msjkm} and near-extremal black hole solutions \cite{cm}
(see also \cite{m1,p}). On the other hand three-dimensional gravity can also
be quantized in a consistent way \cite{at,w1} and therefore one can expect to
find a statistical interpretation
for the BTZ black hole entropy \cite{btz}. Carlip showed \cite{c1} that, by
counting microscopic degrees of freedom of a conformal field theory 
living in an appropriate boundary, one can exactly reproduce the
Bekenstein-Hawking formula for the BTZ black holes. Furthermore, Strominger
\cite{s1} has also been able to obtain the entropy formula by exploiting the
two-dimensional conformal algebra arising as an appropriate symmetry of 
three-dimensional gravity with a negative cosmological constant \cite{bh}.
These results suggest that the statistical explanation of the entropy
is not too much tied to the details of the quantum theory, but rather to general
symmetry properties of the quantum gravity theory. This point of view has
been put forward in \cite{c2,c3,so}.\\

The holographic correspondence between gravity on AdS$_3$ and a two-dimensional
conformal field theory, discovered by Brown and Henneaux, was realized in 
terms of asymptotic symmetries at spatial infinity. This type of realization
of the AdS/CFT correspondence \cite{m2,w2} was analyzed for the
Jackiw-Teitelboim mode of 2D gravity in \cite{cm1,cm2} and further studied in
\cite{nn} in connection with gravity theories around extremal black hole
solutions. The extremal  BTZ and four-dimensional Reissner-Nordstr\"om black
holes possess geometries 
of the form AdS$_2\times$S$^1$ and AdS$_2\times$S$^2$ respectively. 
It was shown in \cite{nn} that the AdS$_2$/CFT$_1$ correspondence, implemented
via asymptotic symmetries, can be used to exactly reproduce the deviation
of the Bekenstein-Hawking entropy from extremality. As it was argued in
\cite{s2}, the symmetry algebra of a one-dimensional conformal field theory is 
just a copy of the Virasoro algebra. The finite-dimensional conformal part
of this Virasoro algebra, the SL(2,${\bf R}$) symmetry, is the isometry group of
anti-de Sitter space in two dimensions. However, we can alternatively regard
the SL(2,${\bf R}$) symmetry as the isometry group of de Sitter space in
two space-time dimensions and consider the Virasoro algebra as its natural
enlargement to the conformal group in one dimension. One of the aims of this
paper is to point out that the realization of the AdS$_2$/CFT$_1$
correspondence in terms of asymptotic symmetries can also be reformulated as 
a dS$_2$/CFT$_1$ correspondence, providing, in turn, a statistical description
of the entropy of Schwarzschild-de Sitter black holes \cite{gh} near the
degenerate solution (i.e. the Nariai solution \cite{na}), which has the
geometry dS$_2\times$S$^2$. This way, the explanation of the entropy for two
physically different situations, near extremal Reissner-Nordstr\"om 
and near degenerate Schwarzschild-de Sitter black holes, is similar and seems
to indicate the universality of the mechanism. The second goal of this paper
is to show that this result is valid in any dimension, thus reinforcing
the idea that the Bekenstein-Hawking entropy can be just derived from
symmetry considerations.\\

The paper is structured as follows. In Sect.2 we review, in a parallel
way, the Reissner-Nordstr\"om and Schwarzschild-de Sitter black hole solutions
and the corresponding degenerate limits: the Robinson-Bertotti
(AdS$_2\times$S$^2$) \cite{ro,be} and Nariai (dS$_2\times$S$^2$) solutions,
respectively. These degenerate solutions represent either black holes
of minimum size (for a given electrical charge) or black holes of maximum
size (for a given cosmological constant $\Lambda >0$). In both cases these
solutions are stable. The degenerate Reissner-Nordstr\"om solution is 
extremal and the Schwarzschild-de Sitter solution possesses two horizons 
(the Schwarzschild black hole horizon and the cosmological one) with the same
size and the same temperature, thus being in thermal equilibrium. In Sect.3
we shall show, also in a parallel way, that the deviation of the
Bekenstein-Hawking entropy of nearly degenerate black holes from the entropy 
of the degenerate solution can be derived, via Cardy's formula \cite{ca},
from the Virasoro algebra of asymptotic symmetries. We shall emphasize the
fact that this mechanism, already introduced in \cite{nn}, works for both
situations: for asymptotic geometries of the form  AdS$_2\times$S$^2$ and also
dS$_2\times$S$^2$. In Sect.4 we shall generalize the above results for 
Reissner-Nordstr\"om and Schwarzschild-de Sitter black holes in any
dimension. Finally, in Sect.5, we state our conclusions.

\setcounter{equation}{0}

\section{Degenerate horizons and (A)dS$_2 \times$S$^2$ geometries}

First of all we shall briefly review the basic facts concerning the emergence
of AdS$_2 \times$S$^2$ and dS$_2 \times$S$^2$ geometries in the near-horizon
limit of Reissner-Nordstr\"om and Schwarz\-schild-de Sitter black holes. The
Reissner-Nordstr\"om (RN) black hole can be described by the metric
\be
\label{em1}
ds^2 = -V(r) dt^2 + V(r)^{-1} dr^2 + r^2 d\Omega^2 \, ,
\ee
where
\be
\label{em2}
V(r) = 1-\frac{2ml^2}{r}+\frac{q^2l^4}{r^2} \, ,
\ee
$q$ is the electrical charge and $l^2=G^{(4)}$ is four-dimensional Newton's
constant. For $m^2 > q^2$, $V(r)$ has two positive
roots corresponding to the inner and external black hole horizons. But in
the limit $m^2 \rightarrow q^2$ the two roots coincide and the horizons
apparently merge. However, this is nothing but an artifact of a poor coordinate
choice. In this degenerate case the Schwarzschild coordinates become
inappropriate since $V(r)\rightarrow0$ between the two horizons. To see what
really happens, let us try
\be
\frac{m^2}{q^2} = 1 + \delta^2 \, ,
\ee
so that the degenerate case is recovered in the limit $\delta
\rightarrow 0$. We can now define new coordinates $\psi$ and $\chi$ by
\be
\label{gpem}
t=\frac{|q|}{\delta}\psi\, , \qquad r=|q|(1+\delta \sin \chi) \, .
\ee
The resultant metric, a first order in $\delta$, is
\be
ds^2 = q^2 \left[ (\cos^2 \chi -\delta \sin \chi \cos 2\chi) d\psi^2 -
(1 +\delta \frac{\sin \chi -\sin 3\chi}{2\cos^2 \chi}) d\chi^2
+ d\Omega^2 \right] \, ,
\ee
and when $\delta=0$ there is a non-trivial geometry between the horizons
\be
\label{rb}
ds^2 = -q^2 (-\cos^2 \chi d\psi^2 + d\chi^2) + q^2 d\Omega^2 \, .
\ee
This is the AdS$_2 \times$S$^2$ Robinson-Bertotti geometry describing the
gravitational field of a covariantly constant electrical field \cite{ro,be}.
The transformation (\ref{gpem}) possesses a remarkable similarity to the
Ginsparg-Perry one \cite{gp} for the degenerate horizon case in the
Schwarzschild-de Sitter (SdS) black hole, where the near-horizon geometry is
the dS$_2 \times$S$^2$ Nariai geometry \cite{na}
\be
\label{nariai}
ds^2 = \Lambda^{-1}(-\sin^2 \chi d\psi^2 + d\chi^2) + \Lambda^{-1} d\Omega^2
\, ,
\ee
and $\Lambda>0$ is the cosmological constant. In both (\ref{rb}) and
(\ref{nariai}) cases, the geometry is given by the product of two
constant-curvature spaces.\\

We shall now rederive the above results in a more general setting. We start
considering the most general spherically symmetrical metric
\be
ds^2 = -A^2(r,t) dt^2 + B^2(r,t) dr^2 + D^2(r,t) d\Omega^2 \, .
\ee
If $D(r,t)\neq$ const. in the above metric, we can perform a coordinate
transformation $r\rightarrow r=D(r,t)$ and, after further coordinate
redefinitions, we can write the above metric in the well known form
\be
\label{bt}
ds^2 = -e^{\tilde{\nu}(r,t)} dt^2 + e^{\tilde{\l}(r,t)} dr^2 + r^2 d\Omega^2
\, .
\ee
The only thing that remains to be done is to impose Einstein's
equations
\be
\label{ee}
G_{\mu \nu} + \Lambda g_{\mu \nu} = 8 \pi l^2 T_{\mu \nu} \, ,
\ee
where $\Lambda$ is the cosmological constant. For a cosmological charged body
($A_{\mu}=(\frac{q}{r},0,0,0)$) the solution (generalized Birkhoff's theorem)
reads as
\be
ds^2 = -U(r;\Lambda,q,m) dt^2 + \frac{dr^2}{U(r;\Lambda,q,m)} + r^2
d\Omega^2 \, ,
\ee
where
\be
U(r;\Lambda,q,m) = 1 - \frac{2ml^2}{r} - \frac{\Lambda}{3} r^2 +
\frac{q^2l^4}{r^2} \, ,
\ee
and $m$ is the mass of the black hole. For $\Lambda=q=0$ we recover the
Schwarzschild black hole, for $\Lambda=0$ the RN black hole and for $q=0$ the 
SdS black hole.\\

It is interesting to comment that, in a different way from the Schwarzschild
black hole, the $\Lambda,q \neq 0$ cases possess a richer physics. Whereas for
the first case the function $U(r;m)$ only has one zero (the black hole
horizon), the presence of new parameters provides more complexity so that the
function $U(r;\Lambda,q,m)$ can have different roots, simple or multiple roots,
depending in which way we adjust the different parameters. One can find some
degenerate cases in which two different horizons become coincident for certain
relations between the parameters $m$, $q$ and $\Lambda$. Two simple examples
of this feature are the RN and SdS black holes. In the
second example there are also two roots (corresponding to the black hole and
the cosmological horizon) for $0<m<\frac{1}{3l^2}\Lambda^{-\frac{1}{2}}$
($\Lambda>0$) that become coincident ($r=\Lambda^{-\frac{1}{2}}$) in the limit
$m\rightarrow \frac{1}{3l^2} \Lambda^{-\frac{1}{2}}$.\\

Now we consider the case\footnote{This case was already noted in \cite{mtw}.}
$D^2(r,t)=r_0^2=$ const. In this case the spacetime decomposes into the product
of a two-dimensional manifold and the two-dimensional spherical surface 
(M$_4=$M$_2\times$S$^2$); M$_2$ with coordinates $t$, $r$, and S$^2$ with
coordinates $\theta$, $\varphi$. We can now proceed in a similar way as the
$D(r,t)\neq$ const. case. By means of some coordinate redefinitions we get the
following metric
\be
ds^2 = -e^{\nu(r,t)} dt^2 + e^{\l(r,t)} dr^2 + r_0^2 d\Omega^2 \, .
\ee
Note that both $D(r,t)\neq$ const. and $D(r,t)=$ const. are different solutions
not being diffeomorphism connected. Thus, the crucial point is to check the
Einstein equations in order to look for possible solutions to the $\nu(r,t)$,
$\l(r,t)$ functions. As in the $D(r,t)\neq$ const. case we immediately obtain
that the above metric should be static. Furthermore it is worthwhile to remark
that these kinds of solutions do not always exist. The simplest example
emerges in the vacuum $T_{\mu \nu}=0$ and with a vanishing cosmological
constant $\Lambda=0$. The non-vanishing components of the Einstein tensor are
\be
\label{et}
G_0^{\phantom{0}0}=G_1^{\phantom{1}1}=-\frac{1}{r_0^2} \, , \qquad
G_2^{\phantom{2}2}=G_3^{\phantom{3}3}=\frac{1}{2} e^{-\l} (\nu^{\prime \prime}
+ \frac{1}{2} \nu^{\prime 2} - \frac{1}{2} \nu^{\prime} \l^{\prime})\, ,
\ee
and it is immediately noticeable that $G_0^{\phantom{0}0}$ and 
$G_1^{\phantom{1}1}$ do not satisfy the vacuum Einstein equations (\ref{ee}).
Instead, if we consider a non-vanishing stress tensor or a cosmological
constant, the situation changes and new solutions for the functions $\nu(r)$
and $\l(r)$ appear. We can get more global information about these solutions
by taking the trace of (\ref{ee}), being the curvature $R=\bar{R}+\hat{R}$,
where
\be
\bar{R} = -e^{-\l} (\nu^{\prime \prime} + \frac{1}{2} \nu^{\prime 2} -
\frac{1}{2} \nu^{\prime} \l^{\prime})\, , \qquad \hat{R} = \frac{2}{r_0^2} \, ,
\ee
are respectively the curvatures of M$_2$ and S$^2$. It follows immediately
that $R=4\Lambda$ and then M$_2$ is also a constant curvature space. Let us
analyze two simple and well-known examples.\\

\noindent \underline{{\bf \# 1.} $T_{\mu}^{\phantom{a}\nu}=0$}\\
In this case the components $G_0^{\phantom{0}0}=G_1^{\phantom{1}1}$ satisfy
the equations (\ref{ee}) for $\Lambda>0$ being the constant $r_0^2=
\Lambda^{-1}$. Thus M$_2$ is a positive constant-curvature space and the
remaining equations solve for the de Sitter space. The global topology is then
dS$_2 \times$S$^2$ and the metric is nothing but the Nariai metric
(\ref{nariai}).\\

\noindent \underline{{\bf \# 2.} $\Lambda=0$}\\
Now $R=0$, $r_0^2=q^{-2}$  and equations (\ref{ee}) solve for a constant stress
tensor. Let us consider the tensor of a constant electrical field
\be
T_0^{\phantom{0}0}=T_1^{\phantom{1}1}=-T_2^{\phantom{2}2}=-T_3^{\phantom{3}3}
=-\frac{1}{8\pi q^2} \, .
\ee
Thus M$_2$ becomes the anti-de Sitter space being the global topology
AdS$_2 \times$S$^2$ and the metric given by (\ref{rb}).\\

In the two above examples the $D(t,r)=$ const. solutions are just the
geometries that we found previously around the degenerate horizon
configurations in the $D(t,r)\neq$ const. solutions. We shall show this in a
more general context in the remaining part of this section. Let us consider
again the general static solution (\ref{bt}) with
$\tilde{\l}(r)=-\tilde{\nu}(r)=\ln U(r;m,\xi)$, where $m$ is the mass and
$\xi$'s are parameters such as the cosmological constant, electrical charge,
etc. The horizons are the roots of $U(r)$. Solutions with horizon degeneracy
will be given by $U(r)$ with two or more roots when two neighbouring roots
become coincident, in say, $r_0$, for some determined relations between the
parameters $m_0=m(\xi)$ as it is shown in Fig. 1.\\

\begin{figure}
\centerline{\epsfig{figure=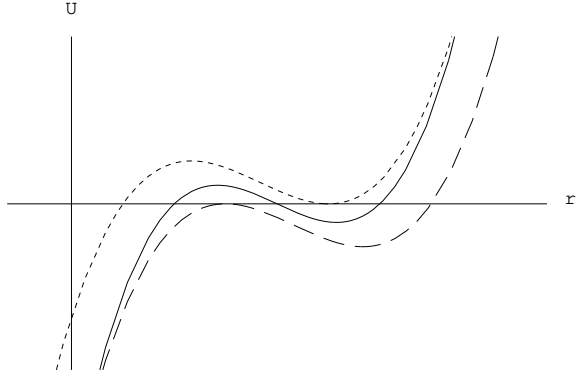,width=3.in}}
\caption{Multi-horizon solutions for different parameter relations. Dotted
and dashed lines are degenerate-horizon configurations.}
\end{figure}

Since $r_0$ is a double root of $U(r;m,\xi)$ for $m=m_0$, it follows
\be
U(r_0;m_0,\xi)=U^{\prime}(r_0;m_0,\xi)=0 \, , \qquad U^{\prime \prime}
(r_0;m_0,\xi)=-\bar{R}_0 \, ,
\ee
where primes denote derivatives with respect to the radial coordinate $r$,
and $\bar{R}_0$ is a constant. Now we perform a perturbative transformation around
the degenerate radius $r_0$ by introducing a new pair of coordinates
$\bar{t}$, $\bar{r}$
\be
t=\frac{\bar{t}}{\a} \, , \qquad r=r_0 + \a \bar{r} \, ,
\ee
where $0<\a \ll 1$. We also write $m=m_0(1+k\a^2)$ where $k$ is an arbitrary
dimensionless constant being positive for $\bar{R}_0<0$ and negative for
$\bar{R}_0>0$. The degenerate case is recovered when $\a=0$. Expanding in
powers of $r-r_0$ in a similar way to what was found in \cite{cfnn},
the metric (\ref{bt}) turns into
\be
\label{nearhm}
ds^2 = -\left( -a^2 -\frac{\bar{R}_0}{2}\bar{r}^2 + \cal{O}(\a)\right)
d\bar{t}^2 + \frac{d\bar{r}^2}{-a^2 -\frac{\bar{R}_0}{2}\bar{r}^2+
\cal{O}(\a)} + \left( r_0^2 + \cal{O}(\a) \right) d\Omega^2 \, ,
\ee
where $a^2=km_0\partial_{m}U(r_0, m, \xi)$,
and still remains a non-trivial geometry in the near-horizon limit $\a
\rightarrow 0$ with constant curvature $R=\bar{R}_0+\frac{2}{r_0^2}$. Note
that $\bar{R}_0$ is positive (negative) depending on the timelike (spacelike)
character of the region between the horizons (see Fig. 1) and, in fact, it can
be written as $\bar{R}_0=\pm \frac{2}{r_0^2}$. Concerning the two examples
considered earlier, we have $r_0=\Lambda^{-\frac{1}{2}}$, $m_0=\frac{1}{3}
\Lambda^{-\frac{1}{2}}$ for the SdS gravity, whereas $r_0=|q|=m_0$ for the EM
gravity.\\

The existence of a connection between the presence of black hole solutions
with horizon degeneracy and (A)dS$_2\times$S$^2$ decomposed solutions
is now clear. The construction of these kinds of solutions from Birkhoff's
theorem is associated with the existence of multi-horizon black hole solutions,
and they also arise as the near-horizon geometries around degenerate horizons.

\setcounter{equation}{0}

\section{Holography and entropy of nearly degenerate RN and SdS black holes}

In this section we shall explain how the deviation of the Bekenstein-Hawking
entropy from extremality for four-dimensional Reissner-Nordstr\"om black holes
can be derived in terms of the asymptotic symmetries of the corresponding
near-horizon geometry. Moreover, we shall also show, in a parallel way, that
this
mechanism can be used to obtain the deviation of the entropy of SdS black holes
from the entropy of the degenerate solution. In both cases the near-horizon
geometry, i.e. the leading order metric in power expansion with respect to the
parameter $\a$, can be written as
\be
ds^2 = -(-a^2 -\frac{\bar{R}_0}{2}\bar{x}^2) d\bar{t}^2 + (-a^2 -
\frac{\bar{R}_0}{2}\bar{x}^2)^{-1} d\bar{x}^2 + r_0^2 d\Omega^2 \, .
\ee
Assuming the following boundary conditions for the asymptotic expansion of the
two-dimensional metric
\bea
g_{\bar{t}\bar{t}} &=& \phantom{-} \frac{\bar{R}_0}{2} \bar{x}^2 +
\g_{\bar{t}\bar{t}} + \ldots \> , \\
g_{\bar{t}\bar{x}} &=& \phantom{-} \frac{\g_{\bar{t}\bar{x}}}{\bar{x}^3} +
\ldots \> , \\
g_{\bar{x}\bar{x}} &=& -\frac{2}{\bar{R}_0} \frac{1}{\bar{x}^2} +
\frac{\g_{\bar{x}\bar{x}}}{\bar{x}^4} + \ldots \> , 
\eea
it is not difficult to see that the infinitesimal diffeomorphisms $\z^a
(\bar{x}, \bar{t})$ preserving the above boundary conditions are
\bea
\z^{\bar{t}} &=& \ep(\bar{t}) - \frac{2}{\bar{R}_0^2\bar{x}^2} \ep''(\bar{t})
+ {\cal{O}}\left( \frac{1}{\bar{x}^4} \right) \, , \\
\z^{\bar{x}} &=& -\bar{x} \ep^{\prime}(\bar{t}) + {\cal{O}}\left(
\frac{1}{\bar{x}} \right) \, ,
\eea
where the prime means derivative with respect to the ``$\bar{t}$'' coordinate,
which is a time-like coordinate for AdS$_2$ ($\bar{R}_0<0$) and space-like for
dS$_2$ ($\bar{R}_0>0$). The ${\cal{O}}\left( \frac{1}{\bar{x}^4} \right)$ terms in
the $\bar{t}$ component are arbitrary and represent the pure gauge
transformations. Choosing for instance
\bea
\z^{\bar{t}} &=& \frac{\a^{\bar{t}(\bar{t})}}{\bar{x}^4} \, , \\
\z^{\bar{x}} &=& \frac{\a^{\bar{x}(\bar{t})}}{\bar{x}}\, ,
\eea
one can show that $\g_{\bar{t}\bar{t}}$, $\g_{\bar{x}\bar{x}}$ and
$\g_{\bar{t}\bar{x}}$ transform as follows
\bea
\d \g_{\bar{t}\bar{t}} &=& -\bar{R}_0 \a^{\bar{x}} \, , \\
\d \g_{\bar{x}\bar{x}} &=& - \frac{8}{\bar{R}_0} \a^{\bar{x}} \, , \\
\d \g_{\bar{t}\bar{x}} &=& \phantom{-} \frac{2}{\bar{R}_0} \a^{\bar{x}} +
2\bar{R}_0 \a^{\bar{t}} \, ,
\eea
and this implies that one can make the gauge choice
\be
\g_{\bar{t}\bar{x}}=0 \, .
\ee
Moreover it is just
\be
\Theta_{\bar{t}\bar{t}}=\kappa \left( \gamma_{\bar{t}\bar{t}}-\frac{1}{2}
\left( \frac{\bar{R}_0}{2} \right)^2 \gamma_{\bar{x}\bar{x}} \right) \, ,
\ee
where $\kappa$ is a constant coefficient, the unique gauge invariant quantity
and it transforms according to the rule
\be
\label{rule}
\delta_{\ep}\Theta_{\bar{t}\bar{t}} = \ep(\bar{t}) \Theta_{\bar{t}\bar{t}}^
{\prime} + 2\Theta_{\bar{t}\bar{t}} \ep^{\prime}(\bar{t}) - \frac{2\kappa}
{\bar{R}_0} \ep^{\prime\prime\prime}(\bar{t}) \, .
\ee
Therefore $\Theta_{\bar{t}\bar{t}}$ behaves as the stress-tensor of a
(one-dimensional) conformal field theory living on the boundary of (A)dS$_2$.\\

A puzzling feature of these two-dimensional geometries, in contrast to the
higher-dimensional anti-de sitter spaces, is the emergence of two disconnected
boundaries at $\bar{x}=\pm\infty$. For AdS$_2$, regarded as part of the
near-horizon geometry of near-extremal Reissner-Nordstr\"om black holes, one of 
the boundaries lies outside the black hole horizon (in the asymptotic flat
region) while the other boundary is inside the horizon. According to the results
of \cite{s1}, where the AdS$_2$
geometry is generated as the near-horizon around extremality of BTZ black holes,
the one-dimensional conformal group is generated by one chiral component
(i.e. one copy of the Virasoro algebra) of the two-dimensional conformal group.
In terms of asymptotic symmetries, this Virasoro algebra lives on the outer
boundary ($x\rightarrow\infty$) and this suggest a similar interpretation for
Reissner-Nordstr\"om black holes (see also \cite{mms}). Moreover, from a general
point of view, if the boundary has several components the Hilbert space of the 
CFT is a tensor product \cite{w3}. In our case this means that the Hilbert
space of the inner boundary should be trivial, without any contribution to the
entropy.
We must also note that the boundaries of dS$_2$ are spacelike and thinking in
terms of the Schwarzschild-de Sitter geometry the relevant boundary 
$x\rightarrow\infty$ lives outside the cosmological horizon (in the
asymptotically de Sitter region). Therefore, the holographic description of the
gravitational degrees of freedom of near-extremal Reissner-Nordstr\"om black
holes are physically different. However,
mathematically we can treat both situations in a similar way. The Fourier
components of the vector fields $\z^a \partial_a$, when $\bar{t}$ is considered
a compact parameter, close down a Virasoro algebra with a vanishing central
charge. Note that for AdS$_2$ it is natural to consider periodicity in the
coordinate ``$\bar{t}$'' while for dS$_2$ the natural periodicity is in the
space-like ``$\bar{t}$'' coordinate. However it is well known that a
canonical realization of these types of asymptotic symmetries is allowed to
have a non-zero central charge. In fact the expression (\ref{rule})) implies
that the Fourier components $L_n^R$ of $\Theta_{\bar{t}\bar{t}}$ (when $0 \leq
\bar{t} \leq 2 \pi \b$) are
\be
\label{mode}
L_n^R = \pm  \frac{1}{2\pi \b} \int_0^{2\pi \b} d\bar{t}
\Theta_{\bar{t}\bar{t}} \b e^{\pm in\frac{\bar{t}}{\b}} \, ,
\ee
where the positive sign is for $\bar{R}_0<0$ (AdS$_2$) and the negative one for
$\bar{R}_0>0$ (dS$_2$), generate a Virasoro algebra
\be
i\{ L_n^R, L_m^R \} \equiv i \d_{\ep_m} L_n^R = (n-m)L_{n+m}^R + 
\frac{c}{12} n^3 \delta_{n,-m} \> ,
\ee
with central charge
\be
c = \mp \frac{24}{\bar{R}_0 \b} \kappa \, , \label{ccc}
\ee
where the positive constant $\kappa$ is a coefficient which should be
determined by the gravitational
effective Lagrangian governing the physics near extremality
or degeneracy.\\

At this point we have to remark that although the values of the central charge
(\ref{ccc}) and $L_0$ depend on the arbitrary parameter $\beta$, the quantity
$cL_0$ is independent of $\beta$. This remark is important since, strictly
speaking, the Cardy formula requires, in general, that $c$ should be the
effective central charge $c_{eff}=c-24\Delta_0$, where $\Delta_0$ is the lowest
eigenvalue of the Virasoro generator $L_0$. 
Since our approach does not offer an explicit construction of the boundary
theory, but rather some general conformal properties of it, we cannot
rigorously determine $c_{eff}$. However, the fact that any physical quantity
should be independent of $\beta$ suggests the equality between $c$ and
$c_{eff}$. A way to implement this is to choose $\beta$ in such a way
that $c_{eff}=1$ and a one-dimensional conformal system which leads to this
effective central charge is that defined by the coadjoint orbits of the Virasoro
group \cite{w4, nn2}.
This system preserves unitarity and leads to the above asymptotic density of
states. In fact its phase space is essentially equivalent to the space of
diffeomorphisms preserving the asymptotic expansion of the metric.\\

We shall now evaluate the corresponding central charges for both classes of
black holes. To this end, and due that the variables ($\bar{t}, \bar{x}$) are
the relevant ones
for the asymptotic symmetries, it is quite useful to reduce the theory
integrating out the angular variables. Let us consider the
Einstein-Maxwell action with a cosmological constant
\be
I^{(4)} = \frac{1}{16\pi G^{(4)}} \int d^4x \sqrt{-g^{(4)}} (R^{(4)} -2\Lambda
+ (F^{(4)})^2) \, .
\ee
Imposing spherical symmetry on the metric
\be
\label{physmetric}
d\tilde{s}_{(4)}^2 = g_{\mu \nu} dx^{\mu} dx^{\nu} + l^2 \psi^2 d\Omega^2 \, ,
\ee
where $l^2=G^{(4)}$, $x^{\mu}$ are the 2D coordinates ($t,x$) and $d\Omega^2$ is
the metric on the two-sphere, and assuming a radial electric field
\be
l^{-2} A_{\mu} = (\frac{q}{r},0,0,0) \, ,
\ee
the above action reduces to
\be
I^{(4)} = \frac{1}{4l^2} \int d^2x \sqrt{-g} l^2 \psi^2 (R + 2|\nabla \psi
|^2 \psi^{-2} + \frac{2}{l^2\psi^2} - 2q^2\psi^{-4} - 2\Lambda) \, ,
\ee
and redefining
\bea
ds^2 &=& \sqrt{\f} d\tilde{s}^2 \, , \\
\f &=& \frac{\psi^2}{4} \, ,
\eea
we arrive at
\be
I^{(4)} = \int d^2x \sqrt{-g} (R\f + l^{-2} V(\f)) \, ,
\ee
where
\be
V(\f) = (4\f)^{-\frac{1}{2}} - l^2q^2(4\phi)^{-\frac{3}{2}} - l^2 \Lambda
(4\f)^{\frac{1}{2}} \, .
\ee
The solutions in terms of the two-dimensional metric $g_{{\mu}{\nu}}$ take the
form
\bea
ds^2 &=& -(J(\f)-lm) dt^2 + (J(\f)-lm)^{-1} dr^2 \, , \\
\f &=& \frac{r}{l} \, ,
\eea
where $J(\f)=\int_0^{\f} d\tilde{\f} V(\tilde{\f})$ and in our case
\be
J(\f) = \frac{1}{2} (4\f)^{\frac{1}{2}} + \frac{1}{2} l^2q^2(4\phi)^
{-\frac{1}{2}} - \frac{1}{6} l^2 \Lambda (4\f)^{\frac{3}{2}} \, .
\ee

The degenerate horizons appear for the zeros $\f_0$ of the potential
($V(\f_0)=0$). If we perturb around the degenerate radius of coincident
horizons
\bea
m &=& m_0(1+k\a^2) \, , \\
t &=& \frac{\tilde{t}}{\a} \, , \\
r &=& r_0 + \a \tilde{x} \, , \\
\f &=& \f_0 + \a \tilde{\f} \, ,
\eea
we have \cite{cfnn}
\be
\label{nearmetric}
ds^2 = -(-\frac{\tilde{R}_0}{2} \tilde{x}^2 - km_0l) d\tilde{t} +
\frac{d\tilde{x}^2}{-\frac{\tilde{R}_0}{2} \tilde{x}^2 - km_0l}
+ \cal{O}(\a) \, ,
\ee
where
\be
\tilde{R}_0 = -\frac{J^{\prime\prime}(\f_0)}{l^2} \, .
\ee
We must stress now that the asymptotic symmetries of the two-dimensional metric
(\ref{nearmetric}) are the same as those of the four-dimensional one since
the $r-t$ part of both metrics only differs by a constant factor
$\sqrt{\f_0}=\frac{\psi_{0}}{2}$. In terms of the two-dimensional
Lagrangian the above expansion reads
\be
I^{(4)} = \a \int d^2x \sqrt{-g} (R\tilde{\f} + l^{-2} V^{\prime}(\f_0)
\tilde{\f}) + {\cal{O}}(\a^2) \, .
\ee
So the leading order is governed by the Jackiw-Teitelboim model
\cite{j} (see also \cite{ckz}). The central charge can be worked out using
canonical methods. The full Hamiltonian ${\cal{H}}$ of the theory, to leading
order in $\a$, is given by
\be
{\cal{H}} = {\cal{H}}_0 + {\cal{K}} \, ,
\ee
where ${\cal{H}}_0$ is the bulk Hamiltonian of the Jackiw-Teitelboim theory
and ${\cal{K}}$ is the boundary term necessary to have well-defined
variational derivatives. Remarkably, the boundary term, after some algebra,
turns out to be proportional to the stress-tensor $\Theta_{\tilde{t}\tilde{t}}$
\be
K(\ep(\tilde{t})) = \ep(\tilde{t})\frac{2\alpha}{l} \bigg(
\gamma_{\tilde{t}\tilde{t}} - \frac{1}{2} \Big( \frac{\tilde{R}_0}{2}
\Big)^2 \gamma_{\tilde{x}\tilde{x}} \bigg) \, ,
\ee 
where the two-dimensional scalar curvature $\tilde{R}_0$ is related to
$\bar{R}_0$ by the expression $\tilde{R}_0=\bar{R}_0(\f_0)^{-\frac{1}{2}}$ and
making use of the identification \cite{fms}
\be
K(\ep(\tilde{t})) = \ep(\tilde{t}) \Theta_{\tilde{t}\tilde{t}} \, ,
\ee
we can determine the coefficient $\kappa$ and hence the central charge, which
then becomes
\be
c = \mp \frac{48\a}{l\tilde{R}_0\b} \, .
\ee
Moreover the value of $L_0^R$ near extremality or degeneracy can also be
calculated without difficulty
\be
L_0^R = \pm m_0 k \a \b \, ,
\ee
and Neveu-Schwartz's generator $L_0^{NS}$ is
\be
L_0^{NS} = L_0^R + \frac{c}{24} \, .
\ee
If $L_0^R \gg c$ the asymptotic density of states given by Cardy's formula is
\be
\label{genericentropy}
\Delta S = 2\pi \sqrt{\frac{cL_0^{NS}}{6}} =
2\pi \sqrt{\frac{8m_0k\a^2}{-\tilde{R}_0 l}} \, .
\ee

Let us now check first that this expression exactly accounts for the deviation
of the near-extremal Bekenstein-Hawking entropy from extremality. For the
Reissner-Nordstr\"om black hole we have
\be
\tilde{R}_0 = R_0(\frac{2l}{r_0}) = -\frac{4}{l^5 |q|^3} \, ,
\ee
and
\bea
c &=& \frac{12 |q|^3 l^4 \a}{\b} \, , \\
L_0^R &=& |q| k \a \b \, , \\
m_0k\a^2 &=& m-m_0 = \Delta m \, .
\eea
So, therefore
\be
\Delta S = 2\pi \sqrt{2 |q|^3 l^4 \Delta m} \, ,
\ee
and, as it was pointed out in \cite{nn}, this is just the leading term in the
Bekenstein-Hawking entropy
\be
S^{BH} = \pi l^2 (|q| + \Delta m + \sqrt{2|q| \Delta m + (\Delta m)^2})^2 \, ,
\ee
from the extremal case
\be
S_e^{BH} = \pi l^2 |q| \, .
\ee

We shall now analyze with more detail the Schwarzschild-de Sitter black hole
near degeneracy. The potential function is given by
\be
V(\f) = \frac{1}{2\sqrt{\f}} - 2l^2\Lambda \sqrt{\f} \, ,
\ee
so $\f_0$ is
\be
\f_0 = \frac{1}{4l^2\Lambda} \, ,
\ee
which corresponds to
\be
r_0 = \frac{1}{\sqrt{\Lambda}} \, .
\ee
The curvature $\tilde{R}_0$ is given by
\be
\tilde{R}_0 = -\frac{V^{\prime}(\f_0)}{l^2} = 4l\Lambda^{\frac{3}{2}} \, ,
\ee
which implies that
\be
c = \frac{12\a}{l^2 \Lambda^{\frac{3}{2}} \b} \, .
\ee
The Cardy formula leads to ($\Delta m=m-m_0<0$)
\be
\label{4se}
\Delta S = 2\pi \sqrt{-\frac{2\Delta m}{\Lambda^{\frac{3}{2}} l^2}} \, ,
\ee
and this is exactly the deviation of the Bekenstein-Hawking entropy from the
degenerate solution. Let us see this explicitly. The entropy
associated with the cosmological and black hole horizons, located at $r_+$
and $r_-$ respectively, is given by
\be
S_{\pm}^{BH} = \frac{\pi r_{\pm}^2}{l^2} \, ,
\ee
where $r_+$, $r_-$ are the two positive roots of the polynomial
\be
\frac{\Lambda}{3}r^3 - r - 2l^2m = 0 \, .
\ee
The solutions are
\bea
r_+ &=& \frac{2}{\sqrt{\Lambda}} \cos \frac{\theta}{3} \, , \\
r_- &=& \frac{2}{\sqrt{\Lambda}} \cos (\frac{\theta}{3}+\frac{4\pi}{3}) \, ,
\eea
where $\cos \theta = -3m\sqrt{\Lambda} l^2$. The degenerate case corresponds to
\bea
m_0 &=& \frac{1}{3\sqrt{\Lambda}l^2} \, , \\
r_0 &=& \frac{1}{\sqrt{\Lambda}} \, ,
\eea
so, if $m \lesssim m_0$
\be
\cos \theta \approx -1 - 3\sqrt{\Lambda} l^2 \Delta m \, ,
\ee
then
\be
r_{\pm} \approx \frac{1}{\sqrt{\Lambda}} \left( 1\pm \sqrt{-2l^2\sqrt{\Lambda}
\Delta m} \right) \, ,
\ee
therefore the deviation from the entropy of the degenerate solution is
\be
|\Delta S_{\pm}^{BH}| = \frac{\pi}{l^2} r_0^2 2\sqrt{\frac{-2l^2\Delta m}
{\sqrt{\Lambda}}} = \frac{2\pi}{l} \sqrt{\frac{-2\Delta m}
{\sqrt{\Lambda^{\frac{3}{2}}}}} \, ,
\ee
which agrees with the statistical entropy (\ref{4se}).\\

To end this section we would like to comment briefly on the euclidean solutions
of the above near degenerate black holes. The degenerate solutions is the
product of two spheres with radius $r=r_0$ and the near degenerate solutions
have conical singularities at the horizons. However the near horizon (AdS$_2$
or dS$_2$) geometries (\ref{nearmetric}) leads, in both cases, to euclidean
geometries $S^2\times S^2$ if the euclidean time has period
\be
\b = 2\pi \sqrt{\frac{-2}{l\Delta m \tilde{R}_0}} \, .
\ee
The inverse $\b^{-1}\equiv \Delta T$ gives rise to the deviation of the
temperature from that of the degenerate solution in accordance, via the second
law of thermodynamics, to the entropy deviation (\ref{genericentropy}). This
unravel a common origin of both systems (AdS$_2$ and dS$_2$) for the deviation
of thermodynamical variables.

\setcounter{equation}{0}

\section{Entropy of near-extremal RN and near-degenerate SdS black holes in any
dimension}

The aim of this section is to generalize the results of section 3 for arbitrary
space-time dimensions. Let us start with the Einstein-Maxwell action with a
positive cosmological constant in $(n+2)$ dimensions
\be
\label{naction}
I^{(n+2)} = \frac{1}{16\pi l^n} \int d^{n+2}x \sqrt{-g^{(n+2)}} \left(
R^{(n+2)} - 2\Lambda + (F^{(n+2)})^2 \right) \, ,
\ee
where $l^n$ is Newton's constant $G^{(n+2)}$. The line element of spherically
symmetric solutions is\footnote{See \cite{mp} for RN solutions.}
\be
\label{nmetric}
ds_{(n+2)}^2 = -U(r) dt^2 + \frac{dr^2}{U(r)} + r^2 d\Omega_{(n)}^2 \, ,
\ee
where
\be
U(r) = 1 - \frac{2l^nm}{r^{n-1}\Gamma_{(n)}} + \frac{l^{2n}q^2}{r^{2(n-1)}
\Delta_{(n)}} - \frac{2\Lambda r^2}{n(n+1)} \, ,
\ee
\be
\Gamma_{(n)} = \frac{n {\cal{V}}^{(n)}}{8\pi} \, , \qquad
\Delta_{(n)} = \frac{n}{2(n-1)} \, ,
\ee
${\cal{V}}^{(n)}$ is the area of the unit S$^n$ sphere
\be
{\cal{V}}^{(n)} = \frac{n \pi^{\frac{n+1}{n}}}{\Gamma(\frac{n+1}{n})} \, ,
\ee
and the electromagnetic field is given by
\be
A_{\r} = \left( \frac{lq}{(\frac{r}{l})^{n-1}}, 0, \ldots, 0 \right) \, ,
\qquad \r=0,1, \ldots , n+1 \, .
\ee

The effective theory of the spherically symmetric sector of (\ref{naction}) can
be obtained by dimensional reduction. Decomposing the metric as follows
\be
\label{decompose}
ds_{(n+2)}^2 = d\hat{s}_{(2)}^2(t,r) + l^2 \psi^2(t,r) d\Omega_{(n)}^2 \, ,
\ee
where $d\Omega_{(n)}^2$ is the metric on the n-sphere, the action
(\ref{naction}) reduces to\footnote{The $q=\Lambda=0$ case was already
considered in \cite{kps}.}
\bea
\label{knaction}
I^{(n+2)} &=& \frac{{\cal{V}}^{(n)}}{16\pi l^n} \int d^2x \sqrt{-\hat{g}}
l^n \psi^n \left(
\hat{R} + n(n-1) |\nabla\psi|^2 \psi^{-2} + \right. \nonumber \\
&\phantom{=}& \left. \frac{n(n-1)}{l^2} \psi^{-2} -
2(n-1)^2 q^2 \psi^{-2n} -2\Lambda \right) \, ,
\eea
and performing a redefinition of $\psi$ and a conformal rescaling of the metric
\bea
\label{cr1}
\frac{n}{8(n-1)} \psi^n &\equiv&  D(\psi) = \f \, , \\
\label{cr2}
ds_{(2)}^2 &=& \Omega^2(\f) d\hat{s}_{(2)}^2 \, ,
\eea
where\footnote{See Appendix A for more details.}
\be
\label{conformal1}
\Omega^2(\f) = \frac{n^2}{8(n-1)} \left( \frac{8(n-1)}{n} \f
\right)^{\frac{n-1}{n}} \, ,
\ee
we can eliminate the kinetic term in the action (\ref{knaction}) and then
\be
\label{2Daction}
I^{(n+2)} = \frac{1}{2G} \int d^2x \sqrt{-g} (R\f + l^{-2} V(\f)) \, ,
\ee
where
\be
G = \frac{n\pi}{(n-1) {\cal{V}}^{(n)}} \, ,
\ee
and the potential $V(\f)$ is given by
\bea
V(\f) &=& (n-1) \left( \frac{8(n-1)}{n} \f \right)^{\frac{-1}{n}} - 
\nonumber \\
&\phantom{=}& (n-1)
\frac{l^2q^2}{\Delta_{(n)}} \left( \frac{8(n-1)}{n} \f \right)^{\frac{1-2n}{n}}
- \frac{2l^2\Lambda}{n} \left( \frac{8(n-1)}{n} \f \right)^{\frac{1}{n}} \, .
\eea
The solutions (\ref{nmetric}) transforms into the following solutions of the
effective theory (\ref{2Daction})
\bea
\label{2metric}
ds_{(2)}^2 &=& -(J(\f)-2Glm) dt^2 -(J(\f)-2Glm)^{-1} dx^2 \, , \\
\f &=& \frac{x}{l} \, ,
\eea
where $J(\f)=\int_0^{\f} d\tilde{\f} V(\tilde{\f})$ reads
\bea
J(\f) &=& \frac{n^2}{8(n-1)} \left( \frac{8(n-1)}{n} \f \right)^{\frac{n-1}{n}}
+ \nonumber \\
&\phantom{=}&
\frac{nl^2q^2}{4} \left( \frac{8(n-1)}{n} \f \right)^{\frac{1-n}{n}} -
\frac{2nl^2\Lambda}{n^2-1} \left( \frac{8(n-1)}{n} \f \right)^{\frac{n+1}{n}}
\, .
\eea
The degenerate solutions appear for the zeros of the potential
\be
V(\f_0) = J^{\prime}(\f_0) = 0 \, ,
\ee
and the two-dimensional geometry around the degenerate horizon has a constant
curvature
\be
\tilde{R}_0 = -\frac{J^{\prime\prime}(\f_0)}{l^2} \, .
\ee
A canonical analysis leads to the central charge
\be
c = \mp \frac{24\a}{lG\tilde{R}_0 \b} \, ,
\ee
and a value of $L_0^R$ given by
\be
L_0^R = \pm m_0 k \a \b \, ,
\ee
where we have assumed a periodicity of $2\pi\b$ in $\tilde{t}$. With the above
values the Cardy formula leads to
\be
\Delta S = 2\pi \sqrt{\frac{4m_0k\a^2}{-\tilde{R}_0lG}} \, ,
\ee
and taking into account that
\be
m_0k\a^2 = m-m_0 = \Delta m \, ,
\ee
we get
\be
\label{nstatistical}
\Delta S = 2\pi \sqrt{\frac{4\Delta m}{-\tilde{R}_0lG}} \, .
\ee
We shall now check explicitly that this expression exactly agrees with the
deviation of the Bekenstein-Hawking entropy 
\label{bhe}
\be
S^{BH} = \frac{{\cal{V}}^{(n)} r^n}{4l^n} \, ,
\ee
of a near-degenerate geometry from the entropy of the degenerate solution
\be
\label{oe}
S_0 = \frac{\Nu^{(n)} r_0^n}{4l^n} \, .
\ee
The deviation is then
\be
\label{deviation}
\Delta S^{BH} = \frac{n \Nu^{(n)} r_0^{n-1}}{4l^n} \left. \frac{\partial r}
{\partial \sqrt{\Delta m}} \right|_{\Delta m=0} \sqrt{\Delta m} + {\cal{O}}
(\Delta m) \, .
\ee

\subsection{Reissner-Nordstr\"om black holes}

The radius of the extremal black hole is the double root of
\be
U(r) = 1 - \frac{16\pi l^nm_0}{n \Nu^{(n)}r^{n-1}} + \frac{2(n-1)}{n}
\frac{l^{2n}q^2}{r^{2(n-1)}} \, ,
\ee
where
\be
m_0 = \frac{n}{4} \sqrt{\frac{n}{2(n-1)}} \frac{q}{G} \, ,
\ee
is the mass for the extremal case. Then the radius reads
\be
r_0^{n-1} = \frac{8\pi l^nm_0}{n\Nu^{(n)}} \, .
\ee
We also get
\be
\f_0 = \frac{n}{8(n-1)} \left( \frac{n}{2(n-1)l^2q^2} \right)^{\frac{-n}
{2(n-1)}} \, .
\ee
Expanding around the extremal radius
\be
\label{radius}
r^{n-1} = \frac{8\pi l^n}{n \Nu^{(n)}} m_0 + \frac{16\pi l^n}{\sqrt{2} n
\Nu^{(n)}} \sqrt{m_0 \Delta m} (1+{\cal{O}}(\Delta m)) \, ,
\ee
the entropy deviation (\ref{deviation}), to leading order in $\sqrt{\Delta m}$,
is
\be
\Delta S^{BH} = 2\pi \sqrt{\frac{2r_0^2m_0\Delta m}{(n-1)^2}} = 2\pi
\sqrt{\frac{n^2}{4(n-1)^3} \left( \frac{2(n-1)l^2q^2}{n}
\right)^{\frac{1+n}{2(n-1)}} \frac{l\Delta m}{G}} \, .
\ee
But this exactly coincides with the statistical entropy (\ref{nstatistical})
since, by a straightforward computation, we have that
\be
-l^2 \tilde{R}_0 = J^{\prime\prime}(\f_0) = \frac{16(n-1)^3}{n^2} \left(
\frac{n}{2(n-1)l^2q^2} \right)^{\frac{1+n}{2(n-1)}} \, .
\ee

\subsection{Schwarzschild-de Sitter black holes}

Now we have
\be
U(r) = 1 - \frac{16\pi l^nm_0}{n \Nu^{(n)}r^{n-1}} - \frac{2\Lambda r^2}
{n(n+1)} \, .
\ee
To get the horizons we study the roots of the following polynomial
\be
P(r) = r^{n-1} - \frac{16\pi l^nm_0}{n \Nu^{(n)}} - \frac{2\Lambda}{n(n+1)}
r^{n+1} \, ,
\ee
and we find that for $0<m<m_0$, where
\be
m_0 = \frac{n\Nu^{(n)}}{8\pi l^n} \left( \frac{n(n-1)}{2\Lambda}
\right)^{\frac{n-1}{2}} \, ,
\ee
there are two positive roots $r_-$, $r_+$ that become a double root $r_0$ in
the limit $m=m_0$
\be
r_0 = \sqrt{\frac{n(n-1)}{2\Lambda}} \, .
\ee
For $m>m_0$ there is no root. The physical picture is a black hole in an
asymptotic de Sitter spacetime. $r_-$ and $r_+$ are respectively the radius of
the black hole and cosmological horizons. The degenerate case in which
both horizons merge at $r_0$ is given for $m=m_0$.\\

Now in order to get the entropy deviation (\ref{deviation}) we expand the
polynomial around the degenerate radius and, taking into account that
$m=m_0+\Delta m$ ($0\ll \Delta m <0$) and $P(r_{\pm})=0$, we get
\be
r_{\pm} - r_0 = \pm \sqrt{\frac{2r_0^2}{(n-1)m_0}} \sqrt{|\Delta m|} \, .
\ee
Then the entropy deviation (\ref{deviation}) reads
\be
\label{sdsdeviation}
|\Delta S_{\pm}^{BH}| = 2\pi \sqrt{\frac{n^2}{4(n-1)^2} \left( \frac{n(n-1)}
{2\Lambda l^2} \right)^{\frac{n+1}{2}} \frac{l|\Delta m|}{G}} \, .
\ee
But now
\be
-l^2 \tilde{R}_0 = J^{\prime\prime}(\f_0) = -\frac{16(n-1)^2}{n^2} \left(
\frac{2\Lambda l^2}{n(n-1)} \right)^{\frac{n+1}{2}} \, ,
\ee
where
\be
\f_0 = \frac{n}{8(n-1)} \left( \frac{n(n-1)}{2\Lambda l^2}
\right)^{\frac{n}{2}} \, ,
\ee
and (\ref{sdsdeviation}) exactly coincides with the statistical entropy
(\ref{nstatistical}).

\section{Conclusions and final remarks}

The goal of this paper is to point out that the deviation of the 
Bekenstein-Hawking entropy of nearly degenerate black holes from the
degenerate solution can be computed, via Cardy's formula, from the conformal
asymptotic symmetry of the geometries (A)dS$_2\times$S$^n$ associated with the
degenerate Reissner-Nordstr\"om and Schwarz\-schild-de Sitter black holes.
Partial results has been obtained in a previous paper \cite{nn} and here
we have generalized them to arbitrary dimensions and also for geometries
with a dS$_2$ factor.
We have to stress that our approach does not determine the boundary theory.
However, we have shown that the asymptotic symmetries allow us to determine
the general properties of the theory. Mainly the product $cL_0$, which turns out
to be related with the Bekenstein-Hawking entropy. Our method offers a unified
treatment of physically different black holes and also suggest that the boundary
theory responsible for the entropy of Schwarzschild-de Sitter black hole should
be thought as a conformal (static) field theory rather than a conformal
quantum mechanics.
We can wonder whether these results can also be further
extended to other types of black holes. According to the analysis of
\cite{nn}, this mechanism to derive the entropy for nearly degenerate
black holes works for a generic two-dimensional dilaton gravity theory.
Therefore we can conclude that our approach can be applied to any
higher-dimensional black hole whose thermodynamics can be effectively described
by the thermodynamics of a two-dimensional dilaton theory. So, for instance, the
string black holes considered in \cite{youm} are natural candidates to further
extend our results.

\section*{Acknowledgements}

This research has been partially supported by the CICYT and DGICYT, Spain.
D. J. Navarro acknowledges the Ministerio de Educaci\'on y Cultura for a FPI
fellowship and the Workshop on Particles, Fields and Strings'99 organizers and
British Columbia University for its hospitality during the initial stages
of this work. P. Navarro acknowledges the Ministerio de Educaci\'on y Cultura
for a FPU fellowship. D. J. Navarro also wishes to thank A. Fabbri for useful
conversations. J. Navarro-Salas would like to thank M. Cadoni and S. Carlip for
correspondence.

\appendix
\renewcommand{\theequation}{A.\arabic{equation}}
\setcounter{equation}{0}
\renewcommand{\thesection}{Appendix A}
\section{Conformal redefinitions and dimensional reduction}

In section 4, a conformal reparametrization (\ref{cr1}), (\ref{cr2}) was used in
order to get the effective two-dimensional theory that describes the geometry
close to the degenerate horizon. We shall now state precisely some technical
aspects of it. Let us rewrite the effective action (\ref{knaction}) in the
form
\be
I = \frac{1}{2G} \int d^2x \sqrt{-\hat{g}} \left( D(\psi) \hat{R} +
H(\psi) |\nabla\psi|^2 + l^{-2} \hat{V}(\psi) \right) \, ,
\ee
where $D(\psi)$ is given by (\ref{cr1}) and
\bea
H(\psi) &=& \frac{n^2}{8} \psi^{n-2} \, , \\
\hat{V}(\psi) &=& \frac{n^2}{8} l^{-2} \psi^{n-2} - \frac{n(n-1)}{4} q^2
\psi^{-n} - \frac{n}{4(n-1)} \Lambda \psi^n \, .
\eea
In order to get (\ref{2Daction}) we perform the conformal redefinition
(\ref{cr2}) where $\f=D(\psi)$ and
\be
\label{phiv}
V(\f) = \frac{\hat{V}(\psi(\f))}{\Omega^2(\f)} \, .
\ee
Finally $\Omega^2(\f)$ can be obtained by means of the the following
differential equation \cite{lmgk}
\be
\frac{1}{2} - \frac{dD}{d\psi} \frac{d\ln\Omega}{d\psi} = 0 \, .
\ee
It is
\be
\label{conformal2}
\Omega^2(\f) = C \left( \frac{8(n-1)}{n} \f
\right)^{\frac{n-1}{n}} \, ,
\ee
where $C$ is an integration constant. The new potential (\ref{phiv}) is then
written
\bea
V(\f) &=& \frac{n^2}{8C} \left( \frac{8(n-1)}{n} \f \right)^{\frac{-1}{n}} - 
\frac{n^2}{8C} \frac{l^2q^2}{\Delta_{(n)}} \left( \frac{8(n-1)}{n} \f
\right)^{\frac{1-2n}{n}} - \\ \nonumber
&\phantom{=}& \frac{n}{8(n-1)} \frac{2l^2\Lambda}{C} \left( \frac{8(n-1)}{n}
\f \right)^{\frac{1}{n}} \, .
\eea
In order to determine the constant C, recall that in (\ref{cr2})
$d\hat{s}_{(2)}^2$ and $ds_{(2)}^2$ are given respectively by
(\ref{decompose}) and (\ref{2metric}). It follows immediately that
\be
J(\f) - 2Glm = \Omega^2(\f) U(r) \, ,
\ee
where
\be
\f = \frac{n}{8(n-1)} \left( \frac{r}{l} \right)^n \, .
\ee
We get
\be
\Omega^2(\f) = \left( \frac{n^2}{8(n-1)} \right)^2 C^{-1} \left(
\frac{8(n-1)}{n} \f \right)^{\frac{n-1}{n}} \, ,
\ee
thus comparing with (\ref{conformal2}) we finally obtain
\be
C = \frac{n^2}{8(n-1)} \, ,
\ee
in agreement with (\ref{conformal1})



\end{document}